\documentstyle[prl,aps,epsf,twocolumn]{revtex}
\begin{document}
\title{Effects of Local Fields on Spontaneous Emission in Dielectric Media}
\author {Michael E. Crenshaw and Charles M. Bowden}
\address{Weapons Sciences Directorate, AMSAM-RD-WS-ST,
Aviation and Missile Research, Development, and Engineering Center\\
U.~S.~Army Aviation and Missile Command, Redstone Arsenal, AL 35898-5248, U.S.A.}
\maketitle
\begin{abstract}
The local-field renormalization of the spontaneous emission rate in a dielectric is explicitly obtained from a fully microscopic quantum-electrodynamical, many-body derivation of Langevin--Bloch operator equations for two-level atoms embedded in an absorptive and dispersive, linear dielectric host.
We find that the dielectric local-field enhancement of the spontaneous emission rate is smaller than indicated by previous studies.
\end{abstract}
\pacs{42.50.Ct, 42.50.Lc, 32.70.Jz, 78.55.-m}
\def \varsigma {\zeta}
\par
In the formative period of nonlinear optics, Bloembergen taught us that the nonlinear optical effects of a dilute collection of atoms that are embedded in a dielectric host are enhanced by local-field effects \cite{BIBloem}.
Now, in the era of quantum optics, researchers are again looking at the interaction of dielectric materials, the radiation field, and resonant atoms.
Central to these investigations are efforts to quantize the electromagnetic field in dielectrics.
One widely used technique is to quantize the macroscopic Maxwell equations in which the classical constitutive relations have been assumed \cite{BIDrummond,BIGlauber,BIMilonni,BIMatloob,BIGruner,BIScheel}.
In the microscopic approach, a generalized Hopfield transformation, based on Fano diagonalization, is used to obtain the coupled polariton modes of the coupled field--oscillator system \cite{BIKnoester,BIHuttner,BIJuz}.
These quantization methods are well-established for dielectrics with negligible absorption and techniques to deal with the special requirements of quantizing the electromagnetic field in absorbing dielectrics are beginning to emerge \cite{BIMatloob,BIGruner,BIScheel,BIHuttner,BIJuz}.
\par
Since Purcell first predicted the alteration of the emission rate of an excited atom due to an optical cavity \cite{BIPurcell}, it has become well known that the observed spontaneous emission rate of an atom depends on its environment.
When the quantized coupled field--dielectric theories are applied to the spontaneous emission of a two-level atom embedded in an absorptionless dielectric, the relation
\begin{equation}
\Gamma_{SE}^{diel}= n \ell^2 \Gamma_0
\label{EQaaa01}
\end{equation}
is obtained \cite{BIGlauber,BIMilonni,BIScheel,BIKnoester,BIJuz}.
Here, $n$ is the linear index of refraction and $\ell$ is the dielectric local-field enhancement factor, $\Gamma_0$ is the vacuum spontaneous emission rate, and $\Gamma_{SE}^{diel}$ is the enhanced spontaneous emission rate in the dielectric.
Both the Lorentz virtual-cavity model $\ell=(n^2+2)/3$ and the Onsanger real-cavity model $\ell=3n^2/(2n^2+1)$ have been utilized in various studies of local-field effects on spontaneous emission.
\par
One of the key features of these approaches of applying the quantization of fields in dielectrics to spontaneous emission is that the oscillators that comprise the dielectric host are assumed to be unaffected by the presence of the embedded atom.
The dielectric medium, as well as the field, is treated as a local condition at the site of a resonant atom such that the atom interacts with an all-pervasive, nonlocal, quantized effective field, the vacuum polariton modes, rather than the local vacuum radiation field modes and the oscillators.
Because the near-dipole--dipole interaction is the fundamental interaction underlying the Lorentz local field, a many-body approach that explicitly deals with the vacuum radiation field modes and the interactions of the atom with the nearby polarizable particles of the host is clearly needed to accurately evaluate the effects of local fields on spontaneous emission in dielectric media.
\par
In this letter, we develop Langevin--Bloch operator equations of motion for a dense collection of two-level atoms embedded in a dielectric host medium.
The Heisenberg picture is used since this provides the most direct correspondence between the operator equations and the Bloch equations.
We begin the development from a fully microscopic many-body viewpoint in which the material is treated as a disordered mixture of two different species of two-level systems and derive the Heisenburg equations of motion for the material and field mode operators.
Adiabatically eliminating the variables associated with the quantized field modes results in coupled equations of motion for the material variables.
We take the harmonic oscillator limit for one species by assuming that its resonance frequency is sufficiently detuned from the primary species that the atoms remain in the ground state.
Adiabatically eliminating the harmonic oscillators results in a Langevin-Bloch formulation for two-level systems embedded in a dielectric host that exhibits local-field renormalization of the fluctuations, the near-dipole--dipole (NDD) interaction of the two-level atoms, the radiation field, the dephasing rate, and the population decay rate.
We obtain
\begin{equation}
\Gamma_{SE}^{diel}= Re(\ell) \Gamma_0
\label{EQaaa02}
\end{equation}
for the renormalized spontaneous emission rate.
In our many-body derivation, the dielectric local-field enhancement factor $\ell=(n^2+2)/3$ arises from the interaction of the embedded atom with the nearby polarizable particles of the host via the electromagnetic field.
The linear index of refraction is complex and frequency dependent and properly accounts for the frequency dispersion and absorption of the dielectric.
\par
We consider a disordered mixture of two species, $a$ and $b$, of two-level systems.
The two-level systems are coupled only via the electromagnetic field.
We allow for the possibility of an externally applied probe or driving field that is taken to be in a coherent state.
The constituents of the total Hamiltonian are the Hamiltonians for the free atoms of species $a$ and $b$, the free quantized radiation field, and the interaction of the two-level systems with the quantized electromagnetic field.
We have, in the electric-dipole and rotating-wave approximations,
$$
H= \sum_j {{\hbar \omega_a}\over{2}}\sigma_3^{j}
+\sum_n {{\hbar \omega_b}\over{2}}\varsigma_3^{n}
+ \hbar \sum_{l,\sigma} \omega_l a_l^{\dagger} a_l
$$
$$
-i\hbar\sum_j \sum_{l,\sigma}\left ( g_l^{j} a_l \sigma_+^{j} e^{i{\vec k}_l \cdot {\vec r}_j} - {g_l^{j}}^* a_l^{\dagger} \sigma_-^{j} e^{-i{\vec k}_l \cdot {\vec r}_j} \right )
$$
$$
-i\hbar\sum_n \sum_{l,\sigma}\left ( h_l^{n} a_l \varsigma_+^{n} e^{i{\vec k}_l \cdot {\vec r}_n} - {h_l^{n}}^* a_l^{\dagger} \varsigma_-^{n} e^{-i{\vec k}_l \cdot {\vec r}_n} \right ) 
$$
$$
-{{i\hbar}\over{2}}
\sum_j
\left (
\Omega_a\sigma_+^je^{-i(\omega_p t - {\vec k}_p\cdot{\vec r}_j)}
-\Omega_a^*\sigma_-^je^{i(\omega_p t - {\vec k}_p\cdot{\vec r}_j)}
\right )
$$
$$
-{{i\hbar}\over{2}}
\sum_n
\left (
\Omega_b\varsigma_+^n e^{-i(\omega_p t - {\vec k}_p\cdot{\vec r}_n)}
-\Omega_b^*\varsigma_-^n e^{i(\omega_p t - {\vec k}_p\cdot{\vec r}_n)}
\right ) ,
$$
where $a_l^{\dagger}$ and $a_l$ are the creation and destruction operators for the field modes and $\omega_{l}$ is the frequency of the field in the mode $l$.
The vacuum dispersion relation is used throughout, {\it e.g.} ${\vec k}_l=\hat k_l\omega_l/c$, where $\hat k_l$ is a unit vector in the direction of ${\vec k}_l$.
For atoms of species $a$,
$\sigma_3^{j}$ is the population inversion operator and $\sigma_{\pm}^{j}$ are the raising and lowering operators for the $j^{th}$ atom,
$g_l^{j}=(2\pi\omega_l/\hbar V)^{1/2} \mu_a\hat{\bf p}_j\cdot\hat {\bf e}_{\vec k_l,\sigma}$ is the coupling between the atom at position ${\vec r}_j$ and the quantized radiation field,
$\hat{\bf p}_j$ is a unit vector in the direction of the dipole moment at ${\vec r}_j$,
$\omega_a$ is the transition frequency,
$\mu_a$ is the dipole moment,
$\Omega_a=\mu_a{\cal E}/\hbar$ is the Rabi rate,
and ${\cal E}$ is the field envelope associated with the coherent field with carrier frequency $\omega_p$.
For species $b$, $\varsigma_3^{n}$, $\varsigma_{\pm}^{n}$, ${h_l^n}$, ${\vec r}_n$, $\hat{\bf p}_n$, $\omega_b$, $\mu_b$, and $\Omega_b=\mu_b {\cal E}/\hbar$ respectively perform the same functions.
Finally, $V$ is the quantization volume, $\hat {\bf e}_{\vec k_l,\sigma}$ is the polarization vector, and $\sigma$ denotes the state of polarization.
\par
The Heisenberg equations of motion for the material and field mode operators
$$
{{d a_l}\over{dt}}=-i\omega_l a_l
+\sum_j {g_l^{j}}^*\sigma_-^{j} e^{-i{\vec k}_l\cdot{\vec r}_j}
+\sum_n {h_l^{n}}^*\varsigma_-^{n} e^{-i{\vec k}_l\cdot{\vec r}_n}
$$
$$
{{d \sigma_-^{j}}\over{dt}}=-i\omega_a\sigma_-^{j} + \sum_{l,\sigma} {g_l^{j}} \sigma_3^{j} a_l e^{i{\vec k}_l\cdot{\vec r}_j}
+{{1}\over{2}} \sigma_3^{j} \Omega_a e^{-i(\omega_p t-{\vec k}_p \cdot {\vec
r}_j) }
$$
$$
{{d \sigma_3^{j}}\over{dt}}=
-2\sum_{l,\sigma} (g_l^{j} \sigma_+^{j}a_l e^{i{\vec k}_l\cdot{\vec r}_j} + {g_l^{j}}^* a_l^{\dagger} \sigma_-^{j} e^{-i{\vec k}_l\cdot{\vec r}_j})
$$
$$
- \sigma_+^{j} \Omega_a e^{-i(\omega_p t-{\vec k}_p \cdot {\vec r}_j)} -\Omega_a^* \sigma_-^{j} e^{i(\omega_p t-{\vec k}_p \cdot {\vec r}_j) }
$$
$$
{{d \varsigma_-^{n}}\over{dt}}=-i\omega_b\varsigma_-^{n} + \sum_{l,\sigma} {h_l^{n}} \varsigma_3^{n}a_l e^{i{\vec k}_l\cdot{\vec r}_n}
+{{1}\over{2}} \varsigma_3^{n} \Omega_b e^{-i(\omega_p t-{\vec k}_p \cdot {\vec r}_n) }
$$
$$
{{d \varsigma_3^{n}}\over{dt}}=
-2\sum_{l,\sigma} (h_l^{n} \varsigma_+^{n}a_l e^{i{\vec k}_l\cdot{\vec r}_n} + {h_l^{n}}^* a_l^{\dagger} \varsigma_-^{n} e^{-i{\vec k}_l\cdot{\vec r}_n})
$$
$$
- \varsigma_+^{n} \Omega_b e^{-i(\omega_p t-{\vec k}_p \cdot {\vec r}_n)} -\Omega_b^* \varsigma_-^{n} e^{i(\omega_p t-{\vec k}_p \cdot {\vec r}_n) }
$$
are derived from $i\hbar (dO/dt)=[O,H]$.
From this point, we adopt normal ordering in which $a_l^{\dagger}$ appears to the left of the atomic operators and $a_l$ appears to the right.
\par
The Heisenberg equations of motion for the material variables are obtained by elimination of the variables associated with the quantized field modes \cite{BIPolder}.
Then,
$$
{{d \sigma_-^{j}}\over{dt}}=-i\omega_a\sigma_-^{j}(t)
+{{\mu_a}\over{2\hbar}}\sigma_3^j(t){\cal E}(t) e^{-i(\omega_p t - {\vec k}_p\cdot{\vec r}_j)}
$$
$$
+ \sum_{l,\sigma}
{g_l^{j}} \sigma_3^{j}(t) a_l(0) e^{-i(\omega_l t-{\vec k}_l\cdot{\vec r}_j)}
$$
$$
+ \sum_{l,\sigma} \int_0^t dt^{\prime} e^{-i\omega_l(t-t^{\prime})} \sum_{i}{g_l^{i}}^*{g_l^{j}}\sigma_3^{j}(t)\sigma_-^{i}(t^{\prime})
 e^{-i{\vec k}_l\cdot({\vec r}_i-{\vec r}_j)}
$$
$$
+\sum_{l,\sigma} 
\int_0^t dt^{\prime} e^{-i\omega_l(t-t^{\prime})} \sum_{m} {h_l^{m}}^*{g_l^{j}}\sigma_3^{j}(t) \varsigma_-^{m} (t^{\prime}) e^{-i{\vec k}_l\cdot({\vec r}_m-{\vec r}_j)}.
$$
Using standard QED methods in the Markovian approximation \cite{BIPolder,BILouisell,BINDD}, this immediately reduces to
$$
{{d \sigma_-^{j}}\over{dt}}= i\Delta_a\sigma_-^{j}
+ {{\mu_a}\over{\hbar}} \sigma_3^{j}\left ({{{\cal E}}\over{2}}+f^+\right )
-i\epsilon_a\sigma_3^{j}\bar\sigma_-
-{{1}\over{2}}\gamma_a\sigma_-^{j} +
$$
\begin{equation}
\sum_{l,\sigma} \hskip -0.04cm \int_0^t \hskip -0.05cm dt^{\prime} e^{-i(\omega_l-\omega_p)(t-t^{\prime})} \hskip -0.02cm \sum_{m} {h_l^{m}}^*\hskip -0.01cm {g_l^{j}} \sigma_3^{j}(t) \varsigma_-^{m} (t^{\prime}) e^{i{\vec k}_l\cdot({\vec r}_j-{\vec r}_m)}
\label{EQaaa03}
\end{equation}
in a frame of reference rotating at $\omega_p$, such that $\Delta_a=\omega_p-\omega_a$.
In the context of the analysis of Ref.\ \cite{BIPolder}, with normal ordering,
$f^+$ is a Langevin force operator arising from fluctuations of the vacuum field,
the dephasing rate $\gamma_a/2$ is half of the population decay rate $\gamma_a=4\omega_p^3|\mu_a|^2/(3c^3\hbar)$ that is asssociated with the self-field under the condition $i=j$,
and $\epsilon_a=4\pi N_a |\mu_a|^2/(3\hbar)$ is the strength of the near-dipole--dipole (NDD) interaction due to the reaction field of all other atoms, $i \ne j$, of species $a$, where $N_a$ is the relevant number density and $\bar\sigma_-$ is a local spatial average of the operator \cite{BINDD,BIMK}.
The last term in Eq.\ (\ref{EQaaa03}) is the contribution of the reaction field from all the atoms of species $b$ to the $j^{th}$ atom of species $a$.
A similar analysis yields
$$
{{d \sigma_3^{j}}\over{dt}}=
2\Bigg [ i\epsilon_a\sigma_+^{j}\bar\sigma_-
-{{\mu_a}\over{2\hbar}} \sigma_+^{j}{\cal E} 
-{{\mu_a}\over{\hbar}} \sigma_+^{j} f^+ -
$$
$$
\sum_{l,\sigma} \hskip -0.06cm \int_0^t \hskip -0.06cm dt^{\prime} e^{-i(\omega_l-\omega_p)(t-t^{\prime})} \hskip -0.05cm \sum_{m} {h_l^{m}}^*\hskip -0.025cm {g_l^{j}} \sigma_+^{j}(t) \varsigma_-^{m} (t^{\prime}) e^{i{\vec k}_l\cdot({\vec r}_j-{\vec r}_m)}
$$
\begin{equation}
+H.c. \Bigg ] -\gamma_a (\sigma_3^{j}+1)
\label{EQaaa04}
\end{equation}
for the equation of motion of the inversion operator.
\par
In the harmonic oscillator limit for species $b$, we have
$$
{{d \varsigma_-^{n}}\over{dt}}= i\Delta_b\varsigma_-^{n}
-{{\mu_b}\over{\hbar}}\left ({{{\cal E}}\over{2}}+f^+\right )
+i\epsilon_b \bar\varsigma_-
-{{1}\over{2}}\gamma_b \varsigma_-^{n} +
$$
\begin{equation}
\sum_{l,\sigma} \int_0^t dt^{\prime} e^{-i(\omega_l-\omega_p)(t-t^{\prime})} \sum_{i} {h_l^{n}} {g_l^{i}}^*\sigma_-^{i} (t^{\prime}) e^{i{\vec k}_l\cdot({\vec r}_n-{\vec r}_i)},
\label{EQaaa05}
\end{equation}
where
$\Delta_b=\omega_p-\omega_b$, $\gamma_b=4\omega_p^3|\mu_b|^2/(3c^3\hbar)$, and $\epsilon_b=4\pi N_b |\mu_b|^2/(3\hbar)$.
Equations (\ref{EQaaa03})--(\ref{EQaaa05}) are coupled operator equations for a material composed of two-level systems and harmonic oscillators.
None of the relevant parameters for the two-level systems, {\it i.e.} frequency, field strength, fluctuations, dephasing rate, population decay rate and NDD, are renormalized by the prescence of the host medium as long as we retain separate equations of motion for the harmonic oscillators.
The next step is to adiabatically eliminate the equations of motion for the harmonic oscillators by substituting the formal integral of Eq.\ (\ref{EQaaa05}) into Eqs.\ (\ref{EQaaa03}) and (\ref{EQaaa04}).
Thus
\par
\def \tp {t^{\prime}}
\def \tpp {t^{\prime\prime}}
\def \tppp {t^{\prime\prime\prime}}
$$
{{d \sigma_-^{j}}\over{dt}}= i\Delta_a\sigma_-^{j}
+ {{\mu_a}\over{\hbar}} \sigma_3^{j} \left ( {{{\cal E}}\over{2}}+f^+ \right )
-i\epsilon_a \sigma_3^{j}\bar \sigma_-
- {{1}\over{2}}\gamma_a \sigma_-^{j}
+ \sum_{l,\sigma} \int_0^t dt^{\prime} e^{-i(\omega_l-\omega_p)(t-t^{\prime})}
\sum_{m} {h_l^{m}}^*{g_l^{j}}
e^{i{\vec k}_l\cdot({\vec r}_m-{\vec r}_j)} 
$$
\begin{equation}
\int_0^{\tp} d\tpp e^{\alpha(\tp-\tpp)}
\Bigg [
{{\mu_b}\over{\hbar}} \sigma_3^{j}(t)\left ({{{\cal E}(\tpp)}\over{2}}+f^+(\tpp) \right )
+ \sum_{l^{\prime},\sigma^{\prime}} \int_0^{\tpp} d\tppp e^{-i(\omega_{l^{\prime}}-\omega_p)(\tpp-\tppp)}
\sum_{i} {h_{l^{\prime}}^{m}} {g_{l^{\prime}}^{i}}^* \sigma_3^{j}(t)\sigma_-^{i} (\tppp) e^{i{\vec k}_{l^{\prime}}\cdot({\vec r}_m-{\vec r}_i)} \Bigg ] ,
\label{EQugly}
\end{equation}
where $\varsigma_+^{m}(0)=0$ and $\alpha=i(\Delta_b+\epsilon_b+i\gamma_b/2)$.
\par
The last term is the part of the reaction field that is due to the prescence of the harmonic oscillators and their subsequent adiabatic elimination.
The large square bracket contains terms that are largely equivalent to all of the original field components, the coherent field, vacuum fluctuations, the self-field, and the reaction field, and will lead to the renormalization of each.
The self-field contribution, $i=j$, in which the atom couples to itself via the linear particles, can be evaluated using the transverse delta function \cite{BIKnoester}. 
The reaction field contribution, the near dipole--dipole interaction of all the $i$ atoms with atom $m$, is of the same form as the NDD interaction of the $i$ atoms with atom $j$ that was studied in Ref. \cite{BINDD} and used in obtaining Eq.\ (\ref{EQaaa03}). We refer to this type of interaction as a many-atom Milonni-Knight problem \cite{BINDD,BIMK}.
The sum in the large square brackets becomes
\begin{equation}
-i{{4\pi}\over{3\hbar}} N_a \mu_a^* \mu_b\sigma_3^{j}(t) \bar \sigma_-(\tpp)
+{{1}\over{2}}{{4\omega_p^3}\over{3c^3\hbar}} \mu_a^* \mu_b \sigma_3^{j}(t) \sigma_-^j (\tpp).
\label{EQt1}
\end{equation}
\par
In the last term of Eq.\ (\ref{EQugly}), note that \cite{BILouisell}
\begin{equation}
\int_0^{\tp} d\tpp e^{\alpha(\tp-\tpp)}f(\tpp) \rightarrow {{-1}\over{\alpha}}f(\tp),
\label{EQpresage}
\end{equation}
since the exponential is strongly peaked near $\tp=\tpp$.
Because ${\vec r}_m \ne {\vec r}_j$, the remaining part of the last term in Eq.\ (\ref{EQugly}) is again the many-atom Milonni-Knight problem.
Then the portion exterior to the large square brackets can be written as
\begin{equation}
-{{4\pi}\over{3\hbar}}{{N_b|\mu_b|^2}\over{\Delta_b+\epsilon_b+i\gamma_b/2}}{{\mu_a}\over{\mu_b^*}} \Big [ ... \Big ]
=-(\ell-1) {{\mu_a}\over{\mu_b^*}}\Big [ ... \Big ],
\label{EQt2}
\end{equation}
where $N_b$ is the number density of oscillators, species $b$,
\begin{equation}
\ell={{n_b^2+2}\over{3}}
\label{EQlor}
\end{equation}
is the complex Lorentz dielectric local-field enhancement factor, and $n_b$ is the complex index of refraction of the host material.
Using (\ref{EQt1}) and (\ref{EQt2}) in Eq.\ (\ref{EQugly}), we obtain
$$
{{d \sigma_-^{j}}\over{dt}}= i\Delta_a\sigma_-^{j} - {{\mu_a}\over{\hbar}} \sigma_3^{j} \left ({{\ell{\cal E}}\over{2}}+\ell f^+\right )
+i\ell \epsilon_a \sigma_3^{j}\bar \sigma_-
$$
\begin{equation}
- {{1}\over{2}}\ell\gamma_a  \sigma_-^{j}.
\label{EQresult1}
\end{equation}
\vskip 1.4in
A similar calculation for the equation of motion of the inversion operator, Eq.\ (\ref{EQaaa04}), yields
$$
{{d \sigma_3^{j}}\over{dt}}=
2\Bigg [ i\ell\epsilon_a\sigma_+^{j}\bar\sigma_-
-{{\mu_a}\over{\hbar}} \sigma_+^{j}\left ( {{\ell{\cal E} }\over{2}} +\ell f^+\right )
$$
\begin{equation}
-{{1}\over{4}}\ell\gamma_a (\sigma_3^{j}+1)
+H.c. \Bigg ]
\label{EQresult2}
\end{equation}
\par
Equations (\ref{EQresult1}) and (\ref{EQresult2}) can be considered as operator Langevin--Bloch equations of motion for a dense collection of two-level atoms embedded in a dielectric medium.
The effect of the adiabatically eliminated damped linear oscillators is contained in the complex Lorentz dielectric enhancement factor $\ell$ that renormalizes the coherent field, the Langevin force operator, the NDD interaction, the dephasing rate and the population decay rate.
We have neglected Cauchy principle parts throughout and there will be local-field enhancement effects from these, as well, {\it e.g.} renormalization of the Lamb shift by $Re(\ell)$.
\par
The results of our fully microscopic many-body QED treatment agree with semiclassical results for local-field enhancement of the coherent field \cite{BIBloem} and the NDD interaction \cite{BINDDdiel}, if expectation values are taken in the limit of classical factorization.
The purely quantum mechanical aspects are the enhancement of the Langevin force operator and the corresponding damping rates.
At resonance, the population decay rate that appears in the Langevin--Bloch equation of motion for the inversion operator is the renormalized spontaneous emission rate
\begin{equation}
\Gamma_{SE}^{diel}=Re(\ell(\omega_a)) \gamma_a (\omega_a) = Re(\ell) \Gamma_0 .
\end{equation}
The spontaneous emission rate is renormalized by $Re(\ell)$.
In addition, there is a level shift, or frequency renormalization, in the amount of $Im(\ell)\gamma_a/2$ due to the product of the imaginary part of $\ell$ with the dephasing rate.
\par
There have been a number of experimental measurements of the spontaneous emission rate of embedded atoms, or atom-like particles, in a dielectric \cite{BIRikken}.
However, these experiments typically involve non-trivial boundary conditions, such as organic ligand cages or nanospheres, that can profoundly affect the observed spontaneous emission rate.
To date, we know of no experimental measurements of the index dependence of the spontaneous emission rate in a bulk dielectric.
The complete theory presented here makes it possible to bring the entire arsenal of laser spectroscopy to bear on the measurement of local-field effects due to a dielectric background.
For example, because even small frequency shifts can be resolved spectroscopically, it might be possible to verify our results by measuring the level shift $Im(\ell)\gamma_a/2$ or the renormalization of the Lamb shift by $Re(\ell)$ as a function of the density of a buffer gas.
\par
Operator equations of motion, including the local-field renormalization of the spontaneous emission rate, for a single atom, or a tenuous collection of atoms, embedded in a dielectric are contained in these, more general, results in the limit $\epsilon_a \rightarrow 0$.
The NDD interaction of densely embedded atoms in a dielectric is included primarily because it provides a very useful theoretical backdrop for evaluating the many-body effects.
The same type of many-body summations that occur in the local-field renormalization of the spontaneous emission rate have been studied previously in the evaluation of local-field effects for dense atoms {\it in vacuo} \cite{BINDD}.
Further, it was our semiclassical derivation of the `anomalous' renormalization of NDD interaction in a dielectric host \cite{BINDDdiel} that indicated a need to examine the problem in its entirety.
Because the NDD interaction and the spontaneous emission rate have the same dependence on the dipole moment and arise in the same way from the elimination of the field operator, one would expect them to have the same renormalization in a dielectric.
We have shown that this is the case.
\par
In conclusion, we obtained the renormalization of the spontaneous emission rate of an atom embedded in a dielectric host.
This result was obtained from a fully microscopic, many-body derivation of Langevin--Bloch operator equations for two-level atoms embedded in an absorptive and dispersive, linear dielectric host.
The dielectric local-field enhancement of the coherent field, the Langevin force operator, the NDD interaction, and the damping rates all stem from the same reaction field that arises from the nearby harmonic oscillators, necessitating the full many-body derivation.
We found that the dielectric enhancement of the spontaneous emission rate is much smaller than indicated by previous studies.
This is an enabling result, paving the way for application of high-index materials to enhance quantum optical effects.

\end{document}